\documentstyle[14pt]{article}

\begin{document}

\title{ The model of the ideal crystal as a criterion for
evaluating of the approximate equations of the liquids.}

\author{{\normalsize\bf Yu.V. Agrafonov, A.G. Balakhchi}\\
Department of Physics\\
Irkutsk University\\
Karl Marx 1, Irkutsk 664003\\
Russia,\\
e.mail: bvz@irk.ru}

\date{}
\thispagestyle{ empty}

\maketitle

\begin{abstract}

It is necessary for the statistical description of collective
effects in liquids to set that or other approximation between
direct and pair correlation functions which describe a
neighboring order. The analytical solution of the generalized
Ornstein-Zernike (OZ) equation was obtained for a system of
particles at $T=0$. It is shown that in this limit case the
neighboring order disappears and the direct correlation function
describes a distant order which is typical for the ideal crystal.
The approximation correctly describing the limit transition liquid-solid at
$T=0$ will have a physical meaning.
\end{abstract}

\section* {Introduction}

\ \ \ The structure and thermodynamics of both dense gases and liquids
is described by a pair correlation function $h_{12}$ considering
the collective effects in dense mediums. Nowadays about 20
approximate integral equations of the Ornstein-Zernike (OZ) type
connecting the direct $C_{12}$ and pair $h_{12}$ correlation
functions are proposed for describing of the above-mentioned one
\cite{AG1,AG2}. The degree of exactness of every such
approximation is impossible to be evaluated. The physical meaning
of their basic approximations has not been cleared up to the end
yet. Thus such a situation suggests a choice of maximum exact
approximations in a comparison with the result achieved at the
numerical experiment. Nevertheless such a comparison doesn't
reveal the physical meaning of approximations.

The OZ equation is to be used for the spatial homogenous and
isotropic mediums as in such a case a two-particle distribution
function (as well as a pair correlation function) depends
only on a mutual distance between the particle centers $r_{12}$,
that makes all the calculations much more simplified.

But there exists a more generalized form of the OZ equation for
the spatial inhomogenious anisotropic mediums \cite{AG3,AG4,AG5}. It
has not been used widely yet for the solution of more exact
problems. One can mention its use only for the description of the
liquid-solid phase transition \cite{AG5,AG6}. This work
describes the use of the Martynov-Sarkisov approximation, which
gives the best agreement with the data of the numerical
experiment for the hard spheres system \cite{AG5}.

In this work we suggest using the generalized OZ equation for the
description of the ideal crystal at $T=0$. The pair correlation
function $h_{12}$ in this limit case takes the meaning of the
Dirak $\delta$ function. As a result, a linear integral equation
for the direct correlation function is achieved, which has a
simple analytical solution. It should fit The results known from
crystallophysics for the ideal crystal. So the limit transition
to the model of the ideal crystal may be considered as a physical
criterion for the evaluation of the exactness for the
approximations used in physics of liquids.

\section {Initial equations}

\ \ \ In works \cite{AG3,AG5,AG7} it was shown that a hierarchy of
Bogolubov- Born-Green-Kirkwood-Yvon (BBGKY) equations may be
modified into a two equations system
\begin {equation}
\omega_{1} = n \int G_{2}S_{12}d(2) + \ln a
\label{eq:form1}
\end{equation}
\begin {equation}
h_{12} = C_{12} + n \int G_{3}C_{13}h_{23}d(3)
\label{eq:form2}
\end{equation}
for two unknown functions: a one-particle distribution function
$G_{1}(\vec{r}_{1}) = \exp(\omega_{1}(\vec{r}_{1}))$ and a
two-particle distribution function
\begin {equation}
G_{ij}(\vec{r}_{i}, \vec{r}_{j}) = G_{i}G_{j}(1+h_{ij}),
\label{eq:form3}
\end{equation}
\begin {equation}
h_{ij} = -1+\exp(\frac{-\Phi_{ij}}{\Theta} +
\Omega_{ij}(\vec{r}_{i}, \vec{r}_{j})),
\label{eq:form4}
\end{equation}
where $n=N/V$ is density, $\Phi_{ij}$-molecular interaction
potential, $\Theta=kT$-temperature, $d(i)=d \vec{r}_{i}$- a volume
integral element of the i particle.

All previous distribution functions are expressed through $G_{1}$
and $G_{12}$. The direct correlation functions $S_{ij}$, $C_{ij}$
in equations (\ref{eq:form1}) and (\ref{eq:form2}) are evidently
expressed through infinite integral sets distinguished from the
products of the pair correlation functions $h_{ij}$
\begin {equation}
S_{ij}=h_{ij}-\Omega{ij}-\frac{1}{2}h_{ij}(\Omega_{ij}+\frac{1}{6}
M_{ij}^{(1)}),
\label{eq:form5}
\end{equation}
\begin {equation}
C_{ij}=h_{ij}-\Omega_{ij}+\frac{1}{2}M_{ij}^{(2)},
\label{eq:form6}
\end{equation}
in which
\begin {equation}
M_{ij}^{(k)}=n^{2}\int \int G_{3}G_{4}h_{i3}h_{i4}h_{34}h_{3j}h_{4j}
d(3)d(4)+... , k = 1,2
\label{eq:form7}
\end{equation}
the so-called bridge-functionals represent infinite sums of
non-reduced diagrams.

For spatial homogeneous and isotropic systems (dense gases and
liquids) at $\omega_{1}=0, G_{1}=1, G_{12}=G_{12}(\vec{r}_{12})$,
equation (\ref{eq:form1}) is a definition of the activity factor
logarithm $\ln a$. Equation (\ref{eq:form2}) describing the
neighboring order in liquids tends to come to a well-known OZ
ratio
\begin {equation}
h_{12}=C_{12}+n \int C_{13}h_{13}d(3),
\label{eq:form8}
\end{equation}
being the basis of the modern theory of liquids and molds
\cite{AG1,AG2}. For making ratio (\ref{eq:form8}) closed it is
necessary to set an approximation connecting the direct
correlation function $C_{12}(\vec{r}_{12})$ with the pair
correlation function $h_{12}(\vec{r}_{12})$.

\section{The ideal crystal}

\ \ \ One particle distribution function has the meaning of local density in laboratory coordinate system. For spatially homogeneous isotropic mediums one particle distribution function equals one because the external field does not exist. For the crystal condition one particle distribution function does not equal one, even if the external field does not exist. To describe such a system it is necessary to introduce the external field according to Bogolubov which fixes the crystal place in the space and then to make it tend to zero. Then one particle distribution function describes the local density distribution in the regard to the fixed coordinate system giving the crystal place as a whole. As an origin of such a coordinate system it is convenient to choose one of lattice particles giving it number zero conditionally. Therefore one particle distribution function $G_{1} = \exp(\omega_{1})$ for the crystal can be represented in the form of the periodic function
\begin {equation}
G_{1}(\vec{r}_{1})=`\sum_{\vec{k}}G_{\vec{k}}\exp(\imath \vec{k} \vec{r}_{1}),
\label{eq:form21}
\end{equation}
For the ideal crystal at $T=0$ every Fourier's component $G_{\vec{k}}$ is similar and $G_{1}$ is a superposition of three-dimension Dirak $\delta$ functions
\begin {equation}
G_{1}(\vec{r}_{1})=\sum_{\vec{r}_{n}}\delta(\vec{r}_{1}-\vec{r}_{n})
\label{eq:form22}
\end{equation}
where the summation is done by crystal lattice nodes $\vec{r}_{n}$.whose coordinates, in their turn, must be defined.

The mutual arrangement of lattice nodes (that is the lattice period and the type of the crystal system) is defined by the potential of molecular interaction. On the other hand, the potential $\Phi_{12}$ enters the pair correlation function in the form of the combination $-\frac{\Phi_{12}}{\Theta}$. Let us consider the kind of the function $h_{12}$ at $T=0$. Let the molecular interaction be described by Lenard-Jones potential
\begin {equation}
\Phi_{12}=4\varepsilon[(\frac{\sigma}{r})^{12}-(\frac{\sigma}{r})^{6}],
\label{eq:form23}
\end{equation}
where $\sigma$ is a characteristic size of a molecule. The potential $\Phi_{12}$ has a minimum at the point $r_{0}=\sigma\sqrt[6]{2}$. Therefor the function $-\frac{\Phi_{12}}{\Theta}$ at $T\rightarrow 0$ has $\delta$ -shaped maximum at this point. That is why one can neglect the thermal potential $\omega_{12}$ describing the neighbouring order. As a result $h_{12}$ is approximated one-dimension Dirak function $\delta(r_{12}-r_{0})$.

As a result at $T=0$ we get the following equation for the straight correlation line with regard to (\ref{eq:form22})
\begin {equation}
C^{(2)}_{12}+n\sum_{\vec{r}_{n}} C^{(2)}_{12}(\vec{r}_{1},\vec{r}_{n})\delta(r_{2n}-r_{0})=\delta(r_{12}-r_{0}).
\label {eq:form24}
\end {equation}
We notice that (\ref{eq:form24}) is the linear equation whose solution is sought in the form
\begin {equation}
C_{12}=\alpha\delta(r_{12}-r_{0}),
\label {eq:form25}
\end {equation}
where $\alpha$ is a coefficient to be defined. Substitution $C_{12}$ in (\ref{eq:form24}) results in the following expression 
\begin {equation}
\delta(r_{12}-r_{0})=\alpha\delta(r_{12}-r_{0})+n\alpha\sum_{\vec{r}_{n}} \delta(\mid\vec{r}_{1}-\vec{r}_{n}\mid-r_{0})\delta(\mid\vec{r}_{2}-\vec{r}_{n}\mid-r_{0}).
\label {eq:form26}
\end {equation}
Particles 1 and 2 are the neighbouring ones. The summation is done by those nodes which are the nearest both for the first particle and for the second one. Let such particles have $N_{0}$ numbers. Using (\ref{eq:form26}) we get
\begin {equation}
\alpha=\frac{1}{1+nN_{0}}, r_{12}=r_{0}, r_{13}=r_{0}, r_{23}=r_{0}
 \label {eq:form27}
\end {equation}
The correlation (\ref{eq:form27}) defines the crystal structure (the lattice period and the type of the crystal system). For determining the crystal system we should calculate the number $N_{0}$.

At $T=0$ the system is densely packed. There are two types of lattices with the dense packing - face centered cubic and hexagonal lattices \cite{AG8}. Their structure is formed by successive layers of densety packed planes. Every particle has twelve neighbouring particles. This structure type is connected with the distant order in the crystal. But in any substance (including the crystal) there is the neighbouring order which is described by the second addend in the right-hand side (\ref{eq:form26}) in this case.

Let us take arbitrarily a couple of the neighbourong particles arranged in the plane $(111)$. It has four particles juxtaposed with it for both types of the lattice: two particles in the plane (111) and one particle in every adjoined plane. If particles 1 and 2 lie in two neighbouring planes the situation is different. For the face centered cubic lattice $N_{0}$ is 4 as before but for the hexagonal lattice $N{0}$ is 3. So $N_{0}=$const for the face centered cubic lattice. As the pair of the neighbouring particles is chosen arbitrarily this indicates that the face centered cubic lattice is realized. It corresponds to the experimental data for crystals of inert gases. However for Lenard-Jones potential the hexagonal structure has much lower energy than the face centered cubic lattice therefor it is this lattice that must be realized. The problem of the crystal system is not solved in this paper yet.

Now we show that solutions (\ref{eq:form22}) and (\ref{eq:form25}) satisfy the equation (\ref{eq:form1}). The direct correlation function $C^{(1)}_{12}$ neglecting two particle correlations $\omega_{12}$ equals
\begin {equation}
C^{(1)}_{12}=C^{(2)}_{12}=\alpha\delta(r_{12}-r_{0}).
 \label {eq:form28}
\end {equation}
Substitution this expression in (\ref{eq:form1}) we shall get
\begin {equation}
\omega_{1}=n\alpha\sum_{\vec{r}_{n}}\delta(\mid\vec{r}_{1}-\vec{r}_{n}\mid -r_{0})+\ln a.
\label {eq:form29}
\end {equation}
Let $a=1$ and doing summation by $\vec{r}_{12}$ as it is done in (\ref{eq:form26})
\begin {equation}
\omega_{1}=n\alpha N_{0}\delta(\mid\vec{r}_{1}-\vec{R}_{0}\mid -r_{0})+\ln a.
\label {eq:form30}
\end {equation}
where $\vec {R}_{0}$ is a radius of the vector-particle placed in the coordinate system origin.
Since coordinates of all the nodes are known now, $\omega_{1}$ is a given function $G_{1}$ can be given in the form (\ref{eq:form22}).

The complicity of problems being solved by the physics of
condensed matter makes one use different physically based
approximations. In particular, in the physics of plasma and
solids an approximation of the self-coordinated field is widely
used.

In the physics of liquids a superpositional approximation is
well-known, which is used to express a triple function of the
distribution through two-particle ones. It is to the
superpositional approximation, that fits to the hypernetted chain
equation, which results from (\ref{eq:form2}) by the neglect of all
the irreducible diagrams. Summing up of such diagrams is
practically unrealizable. Physically it is connected with the
fact that at the limit temperatures it is necessary to take into
account the collective effects, stimulated by a molecular motion.
That is why it is necessary to offer a physical value criterion
of the irreducible diagrams.

A limit transition to the ideal crystal at $T=0$ represent such a
criterion. In this case it is possible to neglect the collective
effects connected with the molecular motion. As a result a limit
ratio for the direct correlation function $C_{12}$, that
effectively includes a contribution of the irreducible diagrams,
is possible to be obtained.

From a great number approximations between $C_{12}$ and $h_{12}$
the most physically valid is that one which fits to the given
limit ratio.

\section*{Conclusion}

\ \ \ The solution of Ornstein-Zernike generalized equation is obtained for the ideal crystal at $T=0$. In this case the pair correlation function $h_{12}$ Degenerates to Dirak $\delta$ function and for the direct correlation function $C_{12}$ the linear integral equation is found which has the analytical solution. By using it the main problem of physics of liquids is solved - establishment of closure between the direct and pair correlation functions.

At the same time there is still a problem for the aggregate condition, the method developed by us can be used as a criterion for evaluating the closures: those resulting in the solution for the ideal crystal at $T=0$ will have the physical meaning.

\end{document}